%% file: V372c.tex
\documentclass[rnote]{aa}  
\usepackage{graphicx}
\usepackage{natbib}
\bibpunct{(}{)}{;}{a}{}{,}

\usepackage{txfonts}
\begin{document}
   \title{Five colour photometry of the RRd star V372~Ser\thanks{Photometric 
     data are only available in electronic
     form at CDS via anonymous ftp to cdsarc.u-strasbg.fr (130.79.128.5)
     or via http://cdsweb.u-strasbg.fr/cgi-bin/qcat?J/A+A/}}

   \author{J. M. Benk\H{o}\thanks{Guest observer at Teide 
 Observatory, IAC, Spain}\inst{}
          \and
          S. Barcza}

   \offprints{S. Barcza}

   \institute{Konkoly Observatory of the Hungarian
              Academy of Sciences, P.O. Box 67, 
              H-1525 Budapest, Hungary\\
              \email{benko@konkoly.hu, barcza@konkoly.hu}
             }

   \date{Received; accepted}

% \abstract{}{}{}{}{} 
% 5 {} token are mandatory
  \abstract
   {}
   {The first 
   $U$-band 
   and new
   $BV(RI)_{\rm C}$ 
   photometry of the RRd star \object{V372 Ser} is presented to
    determine some parameters of the star.}
   {In April, May 2007 2812 
   $U,B,V,R_{\rm C},I_{\rm C}$
   frames were obtained at Konkoly and Teide Observatories,
   1508
   $V$ observations were collected from the literature. 
   Fourier fitted light curves have been derived in all
   bands.}
  %results heading (mandatory)
   {The non-linearly coupled frequencies
    $f_0=(2.121840\pm .000001)~\mbox{cycles $\cdot$ day$^{-1}$}$,
    $f_1=(2.851188\pm .000001)~\mbox{c $\cdot$ d$^{-1}$}$, 
    {i.e. periods
    $P_0=0.4712891\pm .0000002$~days,
    $P_1=0.3507310\pm .0000001$~d,
    $P_1/P_0=0.7441950$,
    amplitudes
    $A_0^{(V)}=0.15399$~mag, 
    $A_1^{(V)}=0.20591$~mag, 
    and phases have been found. 
    $A_1/A_0=1.319\pm .008$
    has been found from averaging the amplitude ratio
    in the different bands i.e. the first overtone 
    is the dominant pulsation mode.} From the
    $V$ observations upper limits are given for secular change
    of the Fourier parameters. The period ratio and
    period put V372 Ser among the RRd stars of the 
    globular clusters M3 and IC 4499,
    mass, luminosity, and metallicity estimates are given.    
   }
   {}


   \maketitle
%
%________________________________________________________________

\section{Introduction}

Double mode pulsation plays a key role in the mass determination 
of RR Lyrae (RRL) variable stars because confirmed RRL type
component is not known in binary systems. We have indirect methods 
to determine their mass
\citep{smit1}: 
for double mode RRL (RRd) stars the Petersen diagram \citep{pete1} 
offers a possibility of most accurate determination. 
Of course, the derived values depend on theoretical input. 

In spite of the importance of RRd stars there are very few observed 
data of their multicolour photometric behaviour.
It is partially due to the moderate number of known stars until the 
near past: at the moment of 
this writing we know 30 more or less bright RRd stars 
($V < 14$~mag) in the Galactic field discovered by the NSVS and 
ASAS all-sky variability surveys \citep{nsvs1,asas1,pile1,khru1}. Only 
two stars (\object{GSC 4868-0831} and V372~Ser) are known with
$V < 11.5$~mag. 
A number of fainter objects were discovered in the Galactic bulge 
\citep{mize1} and several more in extragalactic systems allowing to
draw statistically significant conclusions concerning distribution of
the RRd stars in the Petersen diagram, etc (see \citealt{clem3}
and references therein.) 

To review the literature we found that 
\object{AQ Leo}, \object{CU Com}, 
two fainter Galactic, and nine extragalactic RRd stars were studied
spectroscopically  
\citep{clem1,clem2,grat1}. 
Multicolour photometric 
time series are available for CU~Com \citep[CCD 
$BVI$,][]{clem2},
\object{BS Com} \citep[CCD 
$BVRI$,][]{deka1} and AQ~Leo 
\citep[photoelectric 
$BV$,][]{jerz1,jerz2}.
Some multicolour CCD time series were also 
published for six globular clusters containing at least one RRd star
(\object{IC 4499}, \object{M15}, \object{M68}, \object{M3}, 
\object{NGC 6426}, and \object{NGC 2419}). 

The RRd character of V372~Ser (\object{GSC 5002-0629}, 
$\alpha=15^{\mathrm h}17^{\mathrm m}35\fs03$, 
$\delta=-1\degr5\arcmin17\farcs2$) was
discovered a few years ago \citep{ibvs1}. It is the second brightest
known field RRd star \citep{asas1} and it is especially suited for
high precision CCD photometric observations because in its vicinity 
there are comparison and check stars 
(\object{GSC 5002-0506}, \object{GSC 5002-0525}, \object{GSC 5002-0566})
approximately of the same brightness and colour.

This study presents the first 
$U$-band 
as well as new
$BV(RI)_{\rm C}$
time-series data for this RRd star and the fundamental parameters 
which could be derived from the frequencies.
Sect.~2 describes the observations and reductions,
Sect.~3 presents the frequency analysis and Fourier fitted colour 
curves. In Sect.~4 a short discussion is given and the conclusions
are drawn.

%__________________________________________________________________

\section{Observations and reduction}

\subsection{Observations}

%--------------------------------------------------------------------------
\begin{table}
\caption{Logbook of the observations of V372~Ser.}\label{obslog}
\begin{tabular}{lrcl}
\hline
\noalign{\smallskip}
HJD$-$2\,400\,000 &  No. of frames  & quality$^{\mathrm{a}}$ & Telescope\\
\noalign{\smallskip}
\hline
\noalign{\smallskip}
54217.3745-.5945 & 305 & 2 & RCC\\
54220.4249-.4670 &  55 & 3 & RCC\\
54221.3889-.5685 & 250 & 3 & RCC\\
54222.3767-.5986 & 310 & 2 & RCC\\
54223.3726-.5865 & 300 & 2 & RCC \\
54242.3355-.5162 & 202 & 2 & RCC \\
54244.3418-.3808 &  48 & 3 & RCC \\
54245.3444-.5280 & 223 & 2 & RCC \\
54245.3971-.6290 & 211 & 1 &IAC80\\
54248.3915-.6535 & 165 & 1 &IAC80\\
54249.4283-.6746 & 250 & 1 &IAC80\\
54250.4068-.6728 & 270 & 1 &IAC80\\
54251.4152-.6580 & 224 & 1 &IAC80\\
\noalign{\smallskip}
\hline
\end{tabular}
\begin{list}{}{}
\item[$^{\mathrm{a}}$] Quality of the night is characterized 
by the extinction coefficient 
$k'_V$
in units 
$V$~mag$\cdot\mathrm{(air mass)}^{-1}$: 
1: stable and $k'_V<0.25$, 
2: stable and $0.25\leq k'_V\leq 0.5$,
3: $k'_V>0.5$ or 
   $k'_V<0.5$ and unstable or
   $k'_V>0.5$ and interruptions because of clouds.
\end{list}
\end{table}

%-----------------------------------------------------------------------

The observational material of V372~Ser was collected in the spring of 2007
with two telescopes (see Table~\ref{obslog} for details). 
The 1-m RCC telescope is mounted at Piszk\'estet\H{o} Mountain Station of 
the Konkoly Observatory. It was used with the UV-enhancement coating 
Versarray 1300B camera constructed by Princeton Instruments.
This device contains a back illuminated EEV
CCD36-40 
$1340\times1300$ 
chip that corresponds to a 
$6\farcm6\times6\farcm8$ 
field of view (FOV) with 
$0\farcs303$~pixel$^{-1}$ 
resolution. 
The camera was driven by QPAsO, a Tcl based observing software running on
Linux platforms developed by Konkoly staff.
The IAC80{\footnote{The 0.82m IAC80 Telescope is operated on the island 
Tenerife by the Instituto de Astrofisica de Canarias in the Spanish
Observatorio del Teide}} telescope of Teide Observatory with its standard 
camera was the other telescope. Its basic parameters are: 
EEV UV-coated back illuminated  
2148$\times$2048 
chip, 
$11'\times10\farcm5$ 
FOV with 
$0\farcs305$~pixel$^{-1}$  
resolution. Standard Johnson
$UBV$ 
and Kron-Cousins
$(RI)_{\rm C}$ 
filters were used in the observations. 

The total number of images is 2812 in all colours, 
more than 560 for each filter. The typical exposure times were 
150, 80, 40, 20, 20~s (IAC80), 120-180, 60, 15-40, 5-20, 10-20~s (RCC) for
$U, B, V, R_{\rm C}, I_{\rm C}$,
respectively. 
Each night sky flats (or alternatively dome flats) and some bias 
frames (5-50) were taken as main calibration images.

\subsection{Data reduction \& photometry}

We used the {\sc iraf/ccdred}{\footnote{{\sc iraf} is distributed by the 
NOAO,
operated by the Association of Universities for Research in Astronomy
Inc., under contract with the NSF.}} package for the standard
reduction procedures: bias and flat field correction. Other
corrections (e.g.~dark current, deferred charge, non-linearity) were also
investigated, but they were found to be negligible.

The star brightness was determined by using the aperture
photometry task {\sc daophot/phot} of {\sc Iraf}.  This method
provides robust flux estimates with small errors for isolated stars such
as V372~Ser, its comparison and the check stars.
For a finding chart see Fig.~1 in the paper \citet{ibvs1}.
Basically, we carried out relative photometry using GSC 5002-0506 (star A 
of \citealt{ibvs1}) as
primary comparison star. GSC~5002-0525 and 
GSC~5002-0566 (stars B and C, 
respectively) were the check stars. 
The red star \object{GSC 5002-0560} 
(star D of \citealt{ibvs1}) was excluded from 
our study because its colour indices 
differ from those of the variable 
and the brightness we measured differs significantly from the
value given by \citet{ibvs1}. It might be a long period red variable
of small amplitude. 

\subsection{Transformation into the standard system}

The instrumental magnitudes were transformed into the  
$UBV(RI)_{\rm C}$ Johnson-Cousins system.  
The standard sequence published by \citet{Landolt}
was used as a source of standard magnitudes. Our standard
$UBV(RI)_{\rm C}$
magnitudes for stars A, B, C
were determined on the nights HJD$-$2\,454\,200=23 and 48. Within the error our
values were identical  to those of \citet{ibvs1}, therefore, we used 
the weighted mean (giving half weight 
to our measurements of 2\,454\,223). Table~\ref{t1} gives our accepted 
magnitudes and colour indices.
%----------------------------------------------------------------------
\begin{table*}
\caption{Standard magnitudes and colour indices of the comparison and
check stars}\label{t1}
\begin{tabular}{llllll}
\hline
\noalign{\smallskip}
      &  $V$ & $U-B$ & $B-V$ & $V-R_{\rm C}$ & $V-I_{\rm C}$  \\
\noalign{\smallskip}
\hline
\noalign{\smallskip}
GSC 5002--0506 & $11.500\pm .024$ & $0.021\pm .005$ & 
                          $0.534\pm .004$ & $0.336\pm .001$ & $0.664\pm .002$ \\
GSC 5002--0525 & $13.524\pm .026$ & $0.156\pm .025$ & 
                          $0.655\pm .005$ & $0.384\pm .004$ & $0.764\pm .004$ \\
GSC 5002--0566 & $12.925\pm .025$ & $0.052\pm .017$ & 
                          $0.549\pm .005$ & $0.329\pm .004$ & $0.668\pm .008$ \\

\noalign{\smallskip}
\hline
\end{tabular}
\end{table*}
%-----------------------------------------------------------------------
Each night colour differences 
$u_{\rm A}-u_{\rm B},...,b_{\rm A}-b_{\rm B},...$
were determined from all frames and correlated with
$U_{\rm A}-U_{\rm B},...,B_{\rm A}-B_{\rm B},...$, where
$u,b,...$
are the instrumental magnitudes of stars A, B, C.
Since almost perfect linear relations and
random scatter were found the magnitude differences 
of V372~Ser were transformed
to the standard 
$UBV(RI)_{\rm C}$
magnitudes by the formulae
\begin{eqnarray}
\label{u1}
U_{\ast}&=&U_{\rm A}+1.016(u_{\ast}-u_{\rm A})+0.012, \nonumber   \\
B_{\ast}&=&B_{\rm A}+1.018(b_{\ast}-b_{\rm A})+0.011, \nonumber   \\
V_{\ast}&=&V_{\rm A}+1.005(v_{\ast}-v_{\rm A})+0.001,    \\
R_{\rm C,\ast}&=&R_{\rm C,A}+1.001(r_{\ast}-r_{\rm A})+0.002,  \nonumber  \\
I_{\rm C,\ast}&=&I_{\rm C,A}+0.981(i_{\ast}-i_{\rm A})+0.013,   \nonumber
\end{eqnarray}
where the asterisk indicates V372~Ser.
Between HJD$-$2\,454\,245=0.3971-.5280 we have simultaneous observations
with the telescopes RCC and IAC80. Zero point corrections 
$-0.043,-.025,-0.034,-0.019,+0.004$ were added to the RCC
$U,B,V,(R,I)_{\rm C}$
magnitudes in order to obtain congruence of the RCC and IAC80
light curves.

The typical errors of individual
observations were between $0.01$ \--- $0.02$~mag in
$B,V,(R,I)_{\rm C}$ and $0.03$~mag in $U$.

The construction of our observational database\footnote{Available 
also at the 
webpage {\tt http://www.konkoly.hu/staff/benko/pub.html}.} 
is simple:
the file includes the heliocentric Julian
date (HJD) and magnitudes in the Johnson-Cousins reference
$U,B,V,R_{\rm C},I_{\rm C}$, respectively.
In the last column of the table a quality flag is given, it indicates
the 13 frames as well which were excluded from the analysis because of 
bad atmospheric conditions (e.g. sudden transparency loss exceeding 1.5~mag).

%-------------------------------------------------------------------
   \begin{figure*}
   \centering
   \includegraphics[width=\textwidth]{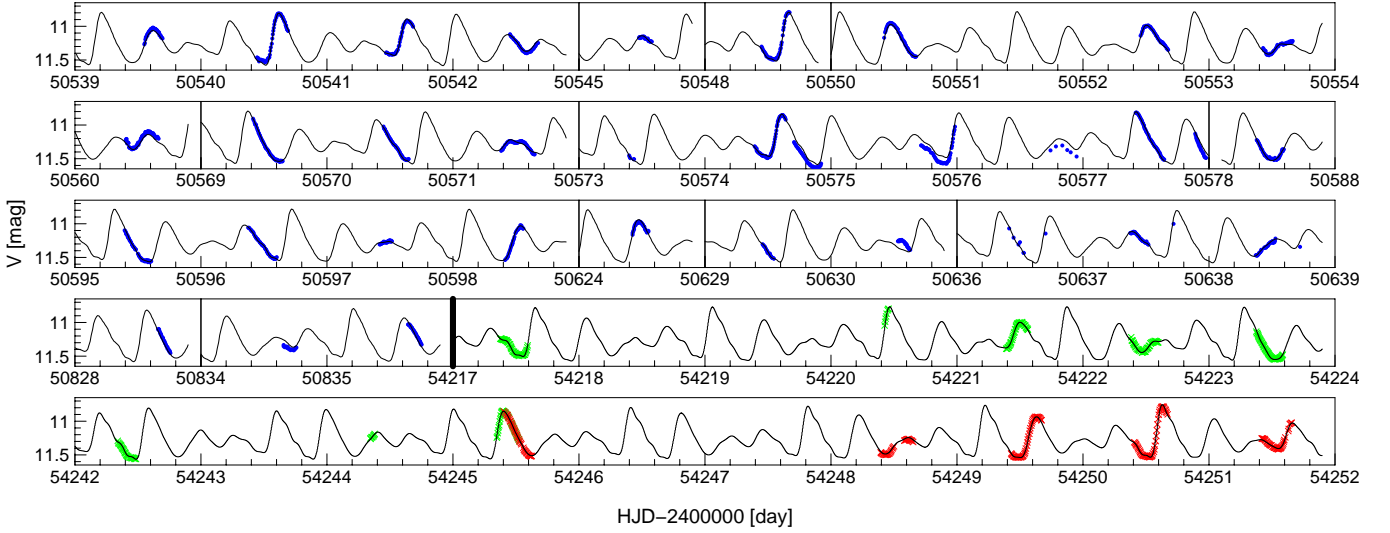}
   \caption{{\it V\/} light curve of V372~Ser. 
            Blue, green, and red symbols denote observations 
	    of \citet{ibvs1}, observations with the 1m RCC 
	    telescope at Konkoly Observatory and  the 
            IAC80 telescope at Teide Observatory, respectively. Line:
            fitted Fourier solution Eq.~(\ref{3.201}).
             }
              \label{fit}%
    \end{figure*}
%-------------------------------------------------------------------

%_____________________________________________________________
%                                             Two column table 
%_____________________________________________________________
%
\begin{table*}
\caption{Fourier decomposition of the {\it UBV(RI)$_{C}$\/} light curves.
Units of frequency, amplitude, phase are 
${\rm d}^{-1}$, mag, and radians, respectively. The first line gives
$A_{00}$ in different bands.
}    
\label{Freq}      
\centering          
\begin{tabular}{@{}*{12}{l}}     % 12 columns 
\hline\hline       
           \noalign{\smallskip}
$\nu_{jk}$ & frequency & $A_{jk}^{(U)}$ & $\varphi_{jk}^{(U)}$ &  
                         $A_{jk}^{(B)}$ & $\varphi_{jk}^{(B)}$ & 
                         $A_{jk}^{(V)}$ & $\varphi_{jk}^{(V)}$ & 
        $A_{jk}^{(R_{\rm C})}$  &  $\varphi_{jk}^{(R_{\rm C})}$ & 
        $A_{jk}^{(I_{\rm C})}$  &  $\varphi_{jk}^{(I_{\rm C})}$ \\
           \noalign{\smallskip}
\hline                  
           \noalign{\smallskip} 
\input{V372_freq.tex}
           \noalign{\smallskip}
\hline                  
\end{tabular}
\end{table*}

\section{Analysis of the light curves}

\subsection{Pulsation frequencies}

We performed standard Fourier frequency analysis on our
time series data using the facilities of the program
package MuFrAn \citep{Muf}.
The frequencies of the fundamental and first overtone
modes ($f_0=2.121841~{\rm d}^{-1}$, $f_1=2.851189~{\rm d}^{-1}$) 
and their $\pm1,2,\dots$ daily aliases
were clearly observed. 
With the first approximation of these two frequencies 
and some harmonics a simultaneous nonlinear fit was carried 
out. Then the light curves were whitened out by subtracting 
the result of the fit. The Fourier spectra of 
the residual light curves contain strong peaks which can be 
identified with the coupling term $f_0+f_1$ and its aliases.
Their appearance is an unambiguous signal of strong nonlinear
coupling in the stellar
pulsation. In order to identify further frequency components,
a successive prewhitening procedure was performed.
The subsequently found frequencies were fitted simultaneously
with all previously identified ones and their overall 
contribution was subtracted from the original data set.
This iterative method increases the signal detection probability
by eliminating the strong alias structure in the vicinity 
of a frequency peak, enabling one to identify hidden low
amplitude components suppressed by aliases and separate 
frequency peaks that are close to each other.  
As a result, 14 frequencies of the light variation were
found (see Table \ref{Freq}).

To determine the final set of frequencies we investigated the CCD
{\it V\/} data set observed in the years 1997-1998 by \citet{ibvs1}.
Performing the above described frequency search
we obtained a Fourier solution with 12 elements of 
very similar parameters to our previous ones, both for the 
frequencies ($f_0=2.12250~{\rm d}^{-1}$, 
$f_1=2.850833~{\rm d}^{-1}$) and the amplitudes. 
Differences between periods are 3.78 and
12.64 seconds for $f_1^{-1}$ and $f_0^{-1}$, respectively: 
the frequency content of the light
variations seems to be unchanged within the last ten years. To verify 
this conjecture we tried to make a common fit of all the  
available {\it V\/} data. 
We remark that because of the lack of data for 10 years
to unify and handle the two data sets in a common
Fourier analysis is not practical due to the 
very unfavourable window function.  
None of the further determined Fourier parameters (frequencies, amplitudes 
and phase) based only either of data sets gave us an acceptable fit for
the unified data. Therefore, we refined the frequency analysis 
by a successive approximation
on the combined data, where the residuals of the non-linear 
fit were minimized while the structure of the decomposition 
(number of harmonics and linear combinations)  was fixed and the two 
frequencies were independently changed
in their limiting intervals. As a result of this process we could fit
the total light curve at the frequencies 
$f_0=2.121840~{\rm d}^{-1}$, $f_1=2.851188~{\rm d}^{-1}$
with the rms accuracy of 0.014 mag.

Since the frequencies were found to be stable we can employ the spectrum
averaging method (SAM) as implemented by \citet{SAM} to find additional
peaks which remained hidden in the individual spectra.
The two spectra were added weighted inversely with 
their variances. The morphology of the peaks of the summed
spectrum confirmed again that the frequencies 
were really constant within their
errors. We subtracted a non-linear fit of the accurate frequencies and their
harmonics from both light curves separately, then the Fourier spectra of
residuals were summed again to find the next frequency and so on.
At the end of this process we achieved a spectrum without any
significant additional peaks at 
$6\sigma$ 
level (for the definition
of significance level of a peak see \citealt{Alc}).
Our conclusion is that additional signal has not been found: 
the Fourier decomposition of our data revealed all significant frequencies.

\subsection{Fourier amplitudes and phase}

\begin{figure}
   \centering
   \includegraphics[width=9cm]{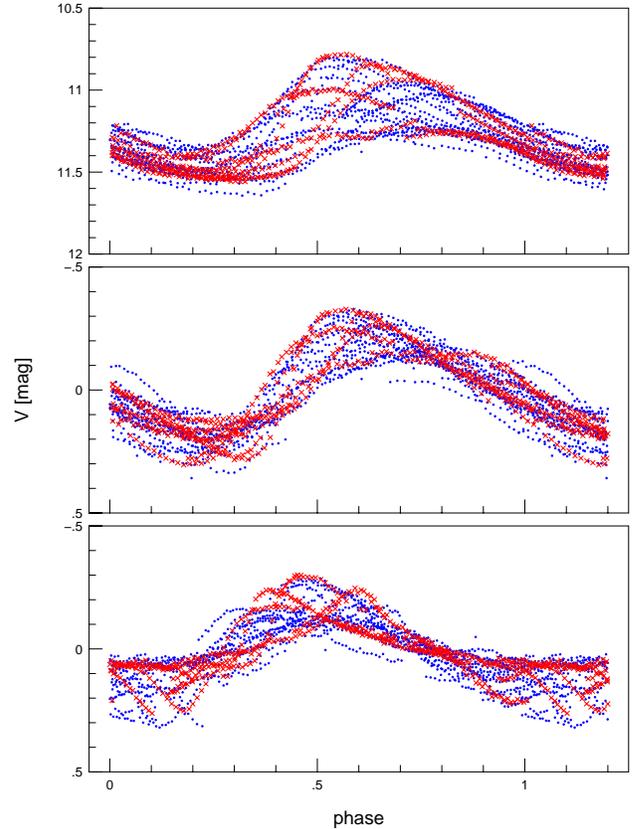}
   \caption{Folded {\it V\/} light curve of V372~Ser, 
            blue and red symbols
            denote observations of \citet{ibvs1} and the
            observations with the 1m RCC and  
            IAC80 telescopes, respectively. 
            Upper panel: folded with
	    $P_1$,
	    middle panel: prewhitened with 
	    $P_0$, folded with
	    $P_1$,
	    lower panel: prewhitened with
	    $P_1$, folded with
	    $P_0$.}
              \label{tekert}%
    \end{figure}

Using the obtained frequencies, their harmonics and
linear combinations non-linear fits were computed determining  
amplitudes and phase for all bands. We assumed 
the light curves $m(t)$ in the form 
   \begin{equation}\label{3.201}
m(t)=A_{00}^{(c)}+\sum_{jk}{A_{jk}}^{(c)}\sin[2\pi\nu_{jk}(t-t_0)
     +\varphi_{jk}^{(c)}],
   \end{equation}
where 
$c=U,B,V,R_{\rm C},I_{\rm C}$,
$\nu_{jk}=jf_0+kf_1$,  
$j; k=0, \pm 1, \pm 2,...$,
and the epoch of an arbitrarily chosen phase was $t_0=2450539$. 
The results are given in Table \ref{Freq}. 
These parameters allowed us to construct 
synthetic light curves. As an example
the synthetic and observed {\it V\/} light curves
are shown in Fig.~\ref{fit}. 
The residual rms scatters of the fits are 0.012, 0.015, 0.015, 0.018 for 
$ B, V, R_{\rm C}, I_{\rm C}$ bands, 
respectively, and it is 
0.028 for {\it U\/}. These values of scatter are compatible with the 
observational accuracy. For the unified {\it V\/} filter 
data (\citealt{ibvs1} and ours) we also found a 
fit with rms=0.014, similar amplitudes, and phase as given in
Table \ref{Freq}. It indicates, that on 10 years scale not only 
frequencies and amplitudes were stable but phase as well. Folded
$V$
light curves are given in Fig~\ref{tekert}.

%-------------------------------------------------------------------
%-------------------------------------------------------------------

The amplitudes of all detected frequencies increase towards 
shorter wavelengths and the amplitudes of the linear combination
frequency components are smaller than those of their constituent
frequencies. All of the detected coupling terms are at positive 
linear combination frequencies of the modes except for $f_0-f_1$.
This seems to be in agreement with model results of \citet{Antonello}.
According to their non-linear transient double-mode Cepheid
models, frequency terms of positive linear combinations
have always larger amplitudes than those of negative linear combination.

To estimate the accuracy of the parameters in Table \ref{Freq}
Monte Carlo simulations were carried out. 1000 artificial time series
were constructed for data in each band 
with the help of the determined Fourier parameters,
Gaussian noise and sampling according to the real observations.
The epochs of the beginning of data sampling were randomly chosen
for the first trial then they were shifted by 0.01 day for each 
subsequently produced synthetic time series. As it was stressed by
\citet{M3} this is important for the interpretation of possible
amplitude variations such as mode changes. The accuracies were found to be
between 0.0025 and 0.0007 mag for all amplitudes. The highest
errors are for the filter {\it U\/} data and the smallest ones are for 
filters 
$ R_{\rm C}$ 
and 
$ I_{\rm C}$. 
Typical errors of the phase are
between 0.01 and 0.004 radians for the two main frequencies. 

\section{Discussion and conclusions}

%--------------------------------------------------------------------------
\begin{table}
\caption{Summary of the relevant data of V372~Ser.}\label{osszefogl}
\begin{tabular}{lc}
\hline
\noalign{\smallskip}
$P_1(\mbox{days})$                        & $0.3507310\pm .0000001$ \\
$P_0(\mbox{days})$                        & $0.4712891\pm .0000002$ \\
$P_1/P_0$                                 & $0.7441950$             \\
$\langle V \rangle=A_{00}^{(V)}$                                 & $11.264$\\
$\langle U-B\rangle=A_{00}^{(U)}-A_{00}^{(B)}$                  & $-0.026$\\
$\langle B-V\rangle=A_{00}^{(B)}-A_{00}^{(V)}$                  & $0.411$ \\
$\langle V-R_{\rm C}\rangle=A_{00}^{(V)}-A_{00}^{(R_{\rm C)}}$  & $0.233$ \\
$\langle V-I_{\rm C}\rangle=A_{00}^{(V)}-A_{00}^{(I_{\rm C)}}$  & $0.443$ \\
$A_{01}^{(V)}$                                                   & $0.2059$\\
$A_{10}^{(V)}$                                                   & $0.1534$\\
$A_{01}^{(U)}/A_{10}^{(U)}$                                     & $1.313$ \\
$A_{01}^{(B)}/A_{10}^{(B)}$                                     & $1.326$ \\
$A_{01}^{(V)}/A_{10}^{(V)}$                                     & $1.337$ \\
$A_{01}^{(R_{\rm C)}}/A_{10}^{(R_{\rm C)}}$                     & $1.299$ \\
$A_{01}^{(I_{\rm C)}}/A_{10}^{(I_{\rm C)}}$                     & $1.318$ \\
\noalign{\smallskip}
\noalign{\smallskip}
\hline
\end{tabular}
\end{table}

%-----------------------------------------------------------------------

Table~\ref{osszefogl} summarizes the relevant data of V372 Ser
derived from our Fourier analysis. The average of the amplitude ratios 
$\overline{A_{0,1}^{(m)}/A_{1,0}^{(m)}}=1.319\pm .008, 
m=U,B,V,R_{\rm C},I_{\rm C}$
shows that the amplitude ratio is independent of colours at
$3\sigma$
level.

From the point of view of frequencies and 
amplitudes V372~Ser is a double mode RR Lyrae star: the period ratio
$0.7441951$
is in the canonical range given by non-linear theoretical values 
expected for RR Lyrae stars \citep{cox1}. The first overtone is the 
dominant mode. In different 
bands identical frequency content has been found:  
the atmosphere is grey from  the point of view of transmitting pulsation 
frequencies.

The magnitude-averaged colour indices 
obtained by fitting the date with Eq. (\ref{3.201}) 
are given in Table~\ref{osszefogl}, they differ slightly from those 
of \citet{ibvs1}: 
$\langle B-V\rangle=0.38,\langle V-R_{\rm C}\rangle=0.26,$
except for 
$\langle V-I_{\rm C}\rangle=0.57$.
We attribute this difference to their much smaller sampling of the
$B, R_{\rm C}, I_{\rm C}$ bands
and to a possible misprint in their
$R_{\rm C}-I_{\rm C}=0.29$
instead of
$0.19$.
V372~Ser has reddish colours even if they are dereddened by the 
upper limit
$E(B-V)=0.08$
for the galactic coordinates of V372~Ser 
\citep{schl1}.

%-------------------------------------------------------------------
   \begin{figure}
   \centering
   \includegraphics[width=9cm]{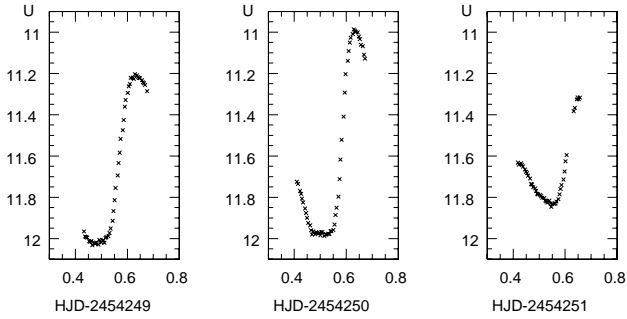}
   \caption{Observed {\it U\/} light curves of a medium, large, and
            small ascending branch}
              \label{ufelszallok}%
    \end{figure}
%-------------------------------------------------------------------

Three
$U$
light curves of night quality 1 are reproduced in Fig.~\ref{ufelszallok}.
They are smooth: hump or bump indicating atmospheric shocks
are not present, in contrast with the ascending branch of e.g. 
\object{SU Dra} \citep{pres1,barc1}. The duration of an ascending branch
is 
$\approx P_1/2$.
 
Upper limits have been estimated for secular change 
${\dot P}=\partial P/\partial t$
of the periods from Monte Carlo simulations and a change
$10^{-6}$ 
in the frequencies in order to find minimal standard deviation of the 
$V$
synthetic light curve and observations:

$\vert{\dot P}_0 \vert \le 6.0\times 10^{-11}\mbox{day/day},$

$\vert{\dot P}_1 \vert \le 2.4\times 10^{-11}\mbox{day/day}$.

\noindent For comparison we mention that 
$\vert{\dot P}\vert < 7\times 10^{-11}$
is typical during the whole lifetime of an RRL star pulsating in single 
mode except for the short phase before leaving the instability strip 
\citep{smit1}. Upper limits for secular amplitude and phase changes are  
$\mbox{O}(\partial A_V/\partial t)\approx 10^{-6}\mbox{mag/day}$,
$\mbox{O}(\partial\varphi_V/\partial t)\approx 10^{-4}\mbox{radians/day}$.

The ratio
$P_1/P_0=0.744195$
and
$P_0=0.471289$
are approximately equal to those of the RRd stars of M3 and
IC 4499, both are moderately metal deficient globular clusters
($[M]=-1.57,-1.50$,
respectively). Interpolating
$P_1/P_0, P_0$
in the Petersen diagram of \citet{bono1} (derived from non-linear, 
non-local, time dependent models with the new opacities) gives  
${\cal M}=0.65\cal{M}_\odot$
and luminosity 
$\log L\approx 1.72$. 
Turbulent convection and more sophisticated hydrodynamic treatment are
added to the above model assumptions in \citep{szab1}. Interpolation in
their diagrams and tables gives 
$[M]\approx -1.7, {\cal M}> 0.7\cal{M}_\odot$
and 
$\log L\approx 1.60$. 

\begin{acknowledgements}
      This work was partially supported by the OTKA Grant K-62304
      and Spanish-Hungarian Bilateral Agreement E-29/04.
      We are grateful to P. \'Abrah\'am and A. Oscoz for calling our
      attention to the vacancy in the time table of telescope IAC80 and
      awarding this to our program on V372~Ser.
      JMB acknowledges the possibility to work at IAC80 telescope
      (Tenerife Observatory, IAC, Spain). We also thank 
      E. Garcia-Melendo for sending us their observational material,
      L. Szabados for reading the manuscript, R. Szab\'o for communicating 
      unpublished material, and an anonymous referee for comments. 
\end{acknowledgements}

\bibliographystyle{aa}

\end{document}

%% file: V372_freq.tex
%freq		freq		A_U		phi_U		A_B		phi_B		A_V		phi_V		A_R		phi_R		A_I		phi_I		\\
-		&-		&11.649		&-		&11.675		&-		&11.264		&-		&11.031		&-		&10.821		&-		\\
$f_0$		&2.121840	&0.20136	&1.41705	&0.19769	&1.33116	&0.15399	&1.28978	&0.12873	&1.22187	&0.09383	&1.15808	\\
$2f_0$		&4.243680	&0.04098	&5.01763	&0.03473	&4.88158	&0.02771	&4.83961	&0.02248	&4.89025	&0.01590	&4.81655	\\
$3f_0$		&6.365520	&0.00823	&2.86820	&0.00816	&3.48286	&0.00828	&3.53757	&0.00588	&3.39453	&0.00541	&3.80429	\\
$f_1$		&2.851188	&0.26444	&0.41696	&0.26206	&0.44289	&0.20591	&0.42275	&0.16723	&0.38875	&0.12366	&0.32015	\\
$2f_1$		&5.702376	&0.04182	&3.98233	&0.04421	&3.99328	&0.03652	&3.98314	&0.02872	&4.01917	&0.02021	&3.93914	\\
$3f_1$		&8.553564	&0.01553	&1.03100	&0.01461	&1.27607	&0.01220	&1.21576	&0.00989	&1.33326	&0.00878	&1.46891	\\
$4f_1$		&11.404752	&0.00542	&4.45459	&0.00729	&4.52769	&0.00632	&4.37625	&0.00525	&4.34081	&0.00360	&4.21666	\\
$f_0+f_1$	&4.973028	&0.09113	&4.09995	&0.08176	&4.16265	&0.06282	&4.16672	&0.05197	&4.16621	&0.03790	&4.14347	\\
$f_1-f_0$	&0.729348	&0.04001	&4.48008	&0.05189	&4.56988	&0.04279	&4.53610	&0.03747	&4.62943	&0.02532	&4.56013	\\
$f_0+2f_1$	&7.824216	&0.02698	&0.88727	&0.02363	&1.01849	&0.01894	&1.04737	&0.01532	&1.11416	&0.01194	&1.08750	\\
$2f_0+f_1$	&7.094868	&0.02190	&1.83531	&0.02122	&2.01262	&0.01600	&1.97154	&0.01470	&2.02126	&0.01064	&1.96231	\\
$2f_0+2f_1$	&9.946056	&0.01630	&5.02404	&0.01961	&5.27763	&0.01453	&5.30904	&0.01267	&5.30237	&0.00855	&5.30350	\\
$f_0+3f_1$	&10.675404	&0.01244	&5.25972	&0.01515	&5.21436	&0.01196	&5.11548	&0.01041	&5.05965	&0.00697	&5.12237	\\
$3f_0+f_1$	&9.216708	&0.00501	&5.39284	&0.01189	&5.45943	&0.01086	&5.51522	&0.00905	&5.63708	&0.00637	&5.68115	\\